\documentclass[conference,compsoc]{IEEEtran}

\ifCLASSOPTIONcompsoc
  \usepackage[nocompress]{cite}
\else
  \usepackage{cite}
\fi

\ifCLASSINFOpdf
\else
\fi

\usepackage{subcaption}

\usepackage[nolist]{acronym}
\usepackage{amsmath}
\usepackage{amssymb}
\usepackage{amsfonts}
\usepackage{gensymb}
\usepackage{cite}
\usepackage{csquotes}
\usepackage{todonotes}
\setuptodonotes{inline}

\newcommand{\banner}[2][]
{\todo[color=gray!10, textcolor=black, #1]{#2}}

\usepackage{hyperref}
\usepackage[binary-units=true]{siunitx}
\usepackage{xspace}
\usepackage{enumitem}
\usepackage[most]{tcolorbox}

\newcommand{\cf}{\textit{cf}.\xspace}
\newcommand{\etal}[0]{\textit{et~al.}\xspace}
\newcommand{\eg}{\textit{e.g.}\xspace}
\newcommand{\ie}{\textit{i.e.}\xspace}

\newtcolorbox{keyobsbox}{
  colback=gray!10,
  colframe=black!60,
  boxrule=0.6pt,
  arc=2pt,
  left=6pt,
  right=6pt,
  top=3pt,
  bottom=3pt,
  fonttitle=\bfseries,
  title=Key Observation,
}

\newtcolorbox{rqbox}[1][]{
  colback=gray!10,
  colframe=black!60,
  boxrule=0.6pt,
  arc=2pt,
  left=6pt,
  right=6pt,
  top=3pt,
  bottom=3pt,
  fonttitle=\bfseries,
  title=Research Question,
}

\newcommand{\Paragraph}[1]{\smallskip\noindent{\bf #1}}

\begin{document}

\title{The Battle of Metasurfaces: Understanding Security in Smart Radio Environments}

\author{\IEEEauthorblockN{  Paul Staat\IEEEauthorrefmark{1}, 
                            Christof Paar\IEEEauthorrefmark{1} and 
                            Swarun Kumar\IEEEauthorrefmark{2}}
        \IEEEauthorblockA{\IEEEauthorrefmark{1}Max Planck Institute for Security and Privacy, Bochum, Germany\\
                          \IEEEauthorrefmark{2}Carnegie Mellon University, Pittsburgh, PA, USA}
        \IEEEauthorblockA{E-Mail: \{paul.staat, christof.paar\}@mpi-sp.org, swarun@cmu.edu}
}

\maketitle

\begin{abstract}
Metasurfaces, or \acp{RIS}, have emerged as a transformative technology for next-generation wireless systems, enabling digitally controlled manipulation of electromagnetic wave propagation. By turning the traditionally passive radio environment into a smart, programmable medium, metasurfaces promise advances in communication and sensing. However, metasurfaces also present a new security frontier: both attackers and defenders can exploit them to alter wireless propagation for their own advantage.
While prior security research has primarily explored unilateral metasurface applications -- empowering either attackers or defenders -- this work investigates \textit{symmetric} scenarios, where both sides possess comparable metasurface capabilities. Using both theoretical modeling and real-world experiments, we analyze how competing metasurfaces interact for diverse objectives, including signal power and sensing perception. Thereby, we present the first systematic study of context-agnostic metasurface-to-metasurface interactions and their implications for wireless security. Our results reveal that the outcome of metasurface “\textit{battles}” depends on an interplay of timing, placement, algorithmic strategy, and hardware scale. Across multiple case studies in Wi-Fi environments, including wireless jamming, channel obfuscation for sensing and communication, and sensing spoofing, we demonstrate that opposing metasurfaces can substantially or fully negate each other’s effects. By undermining previously proposed security and privacy schemes, our findings open new opportunities for designing resilient and high-assurance physical-layer systems in smart radio environments.

\end{abstract}

\IEEEpeerreviewmaketitle

\begin{acronym}

\acro{AP}{access point}

\acro{CSI}{channel state information}

\acro{EM}{electromagnetic}

\acro{FDD}{frequency-division duplex}

\acro{RIS}{reconfigurable intelligent surface}

\acro{LoS}{line of sight}

\acro{MCS}{modulation and coding scheme}

\acro{OFDM}{orthogonal frequency division multiplexing}

\acro{QPSK}{quadrature phase-shift keying}

\acro{PCB}{printed circuit board}

\acro{RF}{radio frequency}
\acro{RSSI}{received signal strength indicator}

\acro{SER}{symbol error rate}
\acro{SNR}{signal-to-noise ratio}
\acro{JSR}{jamming-to-signal ratio}
\acro{SJNR}{signal-to-jamming-and-noise ratio}
\acro{STFT}{short-time Fourier transform}

\acro{SDR}{software-defined radio}

\acro{TDD}{time-division duplex}

\acro{WLAN}{wireless local area network}
\acro{WSN}{wireless sensor network}

\acro{VNA}{vector network analyzer}

\acro{V2X}{vehicle-to-everything}

\acro{MAC}{media access control}

\acro{MIMO}{multiple-input and multiple-output}

\acro{QoS}{quality of service}

\acro{IoT}{Internet of Things}

\end{acronym} 

\section{Introduction}
\label{sec:introduction}

\begin{figure}
    \centering
    \includegraphics[width=0.80\linewidth]{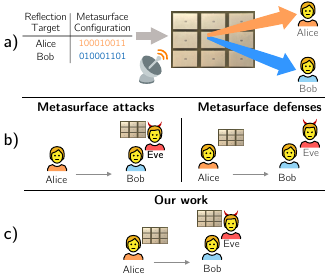}
    \caption{(a)~Illustration of metasurface operation principle. (b)~One-sided metasurface scenarios studied in prior work. (c)~Balanced capability system model studied in this work.}
    \label{fig:system_overview}
\end{figure}

Wireless information systems based on electromagnetic wave propagation are ubiquitous in modern digital society, enabling applications such as smartphones, smart manufacturing, cloud services, connected vehicles, and emergency response systems. Today's widespread wireless adoption is the result of decades of sustained technological innovation to steadily improve performance while reducing overall costs -- a trend that continues as we look towards next-generation wireless systems such as 6G~\cite{jiangRoad6GComprehensive2021}. In this context, one topic that has recently sparked much excitement in wireless research -- both for communication and sensing systems alike -- is metasurfaces, also known as Reconfigurable Intelligent Surfaces {(RIS)}. These are engineered surfaces with digitally tunable reflection properties, realized by combining many individually controllable small reflectors into a larger composite surface (\cf~\autoref{fig:system_overview}a). Metasurfaces are key enablers of the emerging paradigm of \emph{smart radio environments}~\cite{renzoSmartRadioEnvironments2019} where the radio propagation environment is no longer a merely passive and uncontrollable medium, but an active participant in improving wireless performance, alongside the transmitter and receiver. In particular, metasurfaces can engineer signal propagation in a variety of useful ways, \eg, to improve communication signal power, reduce unwanted interference, improve sensing resolution, and impact wireless physical-layer security. Metasurface technology is rapidly moving toward real-world adoption, marked by emerging commercial deployments~\cite{Greenerwave2025, NovoFlectReconfigurableIntelligent2025}, ongoing field trials~\cite{peiRISaidedWirelessCommunications2021}, and active standardization initiatives~\cite{ETSIReconfigurableIntelligent2025}.

The disruptive potential of metasurfaces -- allowing deliberate control over radio propagation that was previously governed solely by natural factors -- has attracted considerable interest in wireless security research. In particular, an increasing number of studies at major security venues examine how metasurfaces can empower either attackers or defenders with new capabilities. For instance, metasurfaces enable both passive and precisely targeted active jamming attacks~\cite{staatMirrorMirrorWall2022, mackensenSpatialDomainWirelessJamming2025}, and can facilitate or mitigate eavesdropping~\cite{shaikhanovMetaFlyWirelessBackhaul2024, weiMetasurfaceenabledSmartWireless2023, liProtegoSecuringWireless2022, xuTwodimensionalHighorderDirectional2024, shaikhanovSpoofingEavesdroppersAudio2025}. In sensing, they allow tailored manipulation to create electromagnetic illusions that protect privacy, evade, or mislead detection systems~\cite{staatIRShieldCountermeasureAdversarial2022, zhouRIStealthPracticalCovert2023, jiangRISirenWirelessSensing2024, shaikhanovSpoofingEavesdroppersAudio2025}. Similarly, metasurfaces can manipulate radar perception to make objects appear or disappear~\cite{reddyvennamMmSpoofResilientSpoofing2023, chenMetaWaveAttackingMmWave2023, wangSpreadSpectrumSelectiveCamouflaging2021, shaoTargetMountedIntelligentReflecting2024}. While prior work has demonstrated the use of metasurfaces to give either attackers or defenders a one-sided advantage over opponents, a natural question emerges: What if both sides employ metasurfaces that interplay?

In this paper, we study a fundamental question about metasurfaces in the security context: \emph{Who prevails in a \enquote{battle} of metasurfaces with conflicting objectives?}  We pose this research question recognizing that metasurfaces are increasingly becoming a baseline capability -- demanding a balanced threat model in which both attackers and defenders have symmetric capabilities. Despite the rich research interest in metasurfaces, this paper, to best of our knowledge, is the first to systematically examine this general, context-agnostic wireless systems question, combining both theoretical analysis and real-world experimental validation. This sets our work apart from prior studies~\cite{alexandropoulosCounteractingEavesdropperAttacks2023, rezvaniLegitimateIllegitimateIRSs2022}, which focused on information-theoretic physical-layer security in scenarios involving both benign and adversarial metasurfaces.

We begin by revisiting prior research on metasurface-assisted wireless jamming attacks~\cite{mackensenSpatialDomainWirelessJamming2025}~(\autoref{sec:jamming_demo}) in a battle scenario as a first case study. Here, we examine how technological parity shifts the threat model from an asymmetric advantage to a strategic competition where a defensive metasurface can completely neutralize the attacker's gain: In our experiments, a metasurface allows the attacker to reduce their jamming power by \SI{6}{\dB} for successful jamming, but a defensive metasurface forces the attacker to increase their signal by \SI{20}{\dB}. Motivated by this observation, we define a simplified game scenario in which two metasurfaces attempt to create constructive and destructive propagation paths between a transmitter and a receiver. This abstraction allows us to explore the interaction between competing metasurfaces while investigating the effect of granting either side precisely controlled advantages. 

The outcome of competing metasurface interactions depends on a range of key factors that determine which side gains an upper hand. First, algorithmic behavior matters, granting the party an advantage who runs better algorithms, is faster, and might possess prior knowledge~(\autoref{sec:factors}). Second, the physical environment defines the upper-bound on the effect of a metasurface, \eg, differences in placement can decisively affect performance (\autoref{sec:channel_contributions}). Third, hardware superiority is important yet not always decisive -- a metasurface with more elements can still underperform due to unfavorable timing or positioning~(\autoref{sec:metasurface_hw}). Finally, destructive objectives dominate constructive ones -- in many cases, the surface attempting to cancel signals has an advantage over one seeking to enhance them (\autoref{sec:metasurface_linearity}). These insights illustrate that dominance in metasurface interactions is determined by an intricate combination of strategy, context, and hardware capabilities.

We evaluate metasurface battle dynamics further through three additional real-world case studies of metasurface-based security and privacy mechanisms in \mbox{Wi-Fi} environments. Across all settings, we find that the introduction of a second opposing metasurface has potential to significantly, in some cases fully, negate the effect of the unilateral metasurface application. We demonstrate that an attacker can fully compromise a metasurface-based secure communication scheme~\cite{liProtegoSecuringWireless2022}~(\autoref{sec:protego}), improving the eavesdropper’s performance from effective random guessing to nearly error-free reception. In the sensing context, we show that a metasurface attacker under certain conditions can mitigate the effect of a metasurface-based channel obfuscation scheme designed to thwart adversarial wireless sensing~\cite{staatIRShieldCountermeasureAdversarial2022}~(\autoref{sec:irshield}). Here, we show that a metasurface helps the attacker to raise detection rates of human motion from \SI{45}{\percent} to \SI{86}{\percent}. Finally, we show that a defensive metasurface can remove the effect of metasurface-based sensing event spoofing~\cite{jiangRISirenWirelessSensing2024}~(\autoref{sec:risiren}).

Taken together, our work contributes a more nuanced understanding of adversarial interactions between metasurfaces, demonstrating that their role in wireless security is shaped not only by their capabilities but also by how -- and when -- they are used. 
In summary, we make the following key contributions:
\begin{itemize}[nosep]

    \item For the first time, we study the prospects of metasurfaces for security and privacy on the physical-layer under technological parity where both attacker and defender employ metasurfaces.
    
    \item We present a comprehensive theoretical analysis, modeling metasurface interactions in a game-like framework.

    \item We empirically validate our findings through real-world case studies of state-of-the-art metasurface security and privacy applications adjusted for technological parity of attacker and defender.    
\end{itemize}

\section{A Primer on Metasurfaces}

Metasurfaces are engineered surfaces designed to manipulate electromagnetic waves in a controlled manner, enabling customized reflection behavior such as non-specular reflections. When this reflectivity is made digitally reconfigurable, metasurfaces become a foundation for creating \textit{smart radio environments} -- an emerging concept in which the propagation medium itself becomes a tunable resource for wireless communication and sensing~\cite{jiangRoad6GComprehensive2021}. By electronically controlling their reflection characteristics, metasurfaces can, for example, redirect radio waves to improve coverage, eliminate blind spots, or enhance energy efficiency~\cite{liaskosNovelCommunicationParadigm2019, liuReconfigurableIntelligentSurfaces2021, basarWirelessCommunicationsReconfigurable2019}.

A typical metasurface consists of a planar array of $L$~identical unit-cell reflector elements printed on a \ac{PCB}, allowing low-cost fabrication~\cite{ranaReviewPaperHardware2023}. The reflection phase of each element can be independently tuned, often using simple digital logic signals from a microcontroller. In the common case of \SI{1}{\bit} control, each element can switch between two phase states, \SI{0}{\degree} and \SI{180}{\degree}, corresponding to the reflection coefficients of $+1$ and $-1$, respectively. This enables the metasurface configuration to be expressed as a binary vector~$c$, which defines the surface' overall reflection pattern.

To achieve a desired reflection behavior between a transmitter and a receiver -- such as maximizing the received signal power -- the metasurface configuration must be adapted to the surrounding radio environment so that the reflected signal components add coherently with direct propagation paths. To tackle the large configuration space of metasurfaces, \eg, $2^L$ possible states for binary-tunable metasurfaces, practical metasurface control often relies on iterative optimization algorithms with feedback from channel measurements~\cite{feng2021optimization, zou2021robust, tewesIRSenabledBreathTracking2022, kainaShapingComplexMicrowave2014, peiRISaidedWirelessCommunications2021}.

\section{Metasurface Battles}

In this section, we outline the threat model and introduce the notion of metasurface battles and the involved entities. Moreover, we outline our experimental setup.

\subsection{Threat Model and Metasurface Battle Notion}

\begin{figure}
    \centering
    \includegraphics[width=0.78\linewidth]{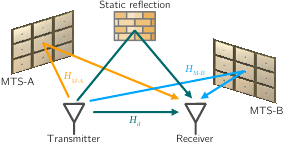}
    \caption{Wireless channel composition in presence of two metasurfaces.}
    \label{fig:system_model}
\end{figure}

We consider a setting with two legitimate wireless devices, named Alice and Bob, that communicate or perform sensing over a single-antenna wireless link. Both devices can estimate the wireless channel, for instance using standard communication preambles to obtain measurements such as \ac{RSSI} or \ac{CSI}. The legitimate parties control a programmable metasurface, denoted as MTS-A, that they configure to redirect or modify radio propagation within the environment. An adversary, Eve, is located in the same environment and possesses radio capabilities equivalent to the legitimate parties. Eve can receive legitimate transmissions to estimate the corresponding wireless channels and can actively transmit their own signals. In addition, Eve controls a second metasurface, denoted as MTS-B, that they reconfigure independently of MTS-A.%

Alice and Bob employ MTS-A to shape the wireless propagation environment to their advantage over Eve, while Eve uses MTS-B to conduct physical-layer attacks against the legitimate parties. We refer to this setting as a \emph{metasurface battle} -- an application-agnostic scenario in which two or more metasurfaces pursue conflicting objectives in controlling the wireless propagation environment, \eg, to alter communication or sensing outcomes. The metasurface-controlled wireless channel thus is the central asset which serves both as the foundation of legitimate applications and as the attack surface subject to adversarial influence. To capture the combined impact of two metasurfaces on the effective wireless channel $H_{\text{eff}}$, we adopt the following model, illustrated in \autoref{fig:system_model}:
\begin{equation}
H_{\text{eff}} = H_d + H_{MA} + H_{MB} + H_{MA,MB}, \label{eq:system_model_basic}
\end{equation}
where all terms represent complex-valued channel coefficients. Here, $H_d$ denotes the direct (non-metasurface) propagation paths, while $H_{MA}$ and $H_{MB}$ correspond to the contributions from the respective metasurface-defined paths. Please note that for the sake of simplicity, we omit potential coupling between MTS-A and MTS-B in \autoref{fig:system_model}, captured by $H_{MA,MB}$. We address mutual metasurface coupling in~\autoref{sec:metasurface_linearity}.

These metasurface channels arise from the superposition of $L$ individual sub-paths between the transmitter, each metasurface element, and the receiver. The channel contributions of metasurfaces~A and~B can be expressed as
\begin{equation}
    H_{Mm} = \sum_{l=1}^{L_m} h^{m}_l c^m_l g^{m}_l \label{eq:mts_subchannels}
\end{equation}
where $h^{m}_l$ and $g^{m}_l$ denote the complex-valued channel coefficients between the transmitter and element~$l$, and between element~$l$ and the receiver, respectively, while $c^m_l$ represents the controllable reflection coefficient of element~$l$, with $m \in \{A, B\}$.
This representation makes explicit that the effective metasurface response is jointly determined by the uncontrollable propagation environment and the tunable reflection states of the individual elements.

\Paragraph{What Constitutes a Win in a Battle?}
A metasurface battle can be viewed as a competition between two metasurfaces, MTS-A and MTS-B, each seeking to shape the wireless environment to meet its own objective. The outcome of this interaction is determined by an \emph{application-specific evaluation function}~$\mathcal{G}$, which quantifies how well the resulting effective wireless environment aligns with each party's objective. The metasurface that achieves a configuration of~$\mathcal{G}$ that is best aligned with its objective is deemed the winner.

At the beginning of a battle, denoted as time~$t_b = 0$, the environment is assumed to be in a neutral or unoptimized state—either without metasurfaces present or with both MTS-A and MTS-B configured randomly such that no meaningful performance gains are realized. As the battle progresses ($t_b > 0$), both metasurfaces adapt their reflection states in an attempt to influence the overall channel. The effectiveness of these actions is evaluated through~$\mathcal{G}$, which can be defined over various measurement domains. Depending on the particular battle,~$\mathcal{G}$ can take different inputs, \eg, simple scalar metrics such as received signal power, or higher-dimensional structures such as time-frequency or spatial channel matrices. Moreover, $\mathcal{G}$ might evaluates multiple different channels at once, \eg, the channel between legitimate parties and that of an eavesdropper.

A simple yet illustrative case arises in a zero-sum setting where two opposing entities Alice and Eve compete over the same effective channel~$H_{\text{eff}}$. Here, Eve seeks to maximize received signal power while Alice aims to minimize it. In this scenario,~$\mathcal{G}$ can be defined as the received signal power (proportional to~$|H_{\text{eff}}|^2$) relative to the initial non-battle state. An increase in Alice’s received power directly corresponds to a degradation in Eve’s, and vice versa. Beyond binary win–loss outcomes, a battle’s result may further be characterized by the \emph{magnitude of gain} achieved by the prevailing metasurface.

\subsection{Experimental Setup}

To study real-world metasurface battles, we use the experimental setup outlined below.

\subsubsection{Metasurface Prototypes}

For MTS-A and MTS-B, we employ two identical metasurface devices based on the open-source design of %
Heinrichs~\etal~\cite{heinrichsOpenSourceReconfigurable2023}.
Each metasurface consists of an FR4-based \ac{PCB} comprising $L = 256$ unit-cell reflector elements arranged in a $16 \times 16$ grid, measuring \SI{36.0}{\cm}~$\times$~\SI{24.7}{\cm}. The unit cells provide binary phase control optimized for operation in the \SI{5}{\GHz} \mbox{Wi-Fi} band. The metasurface is programmable via USB serial communication, allowing the phase of each element $c_l$ to be configured to either \SI{0}{\degree} (state~`\texttt{0}') or \SI{180}{\degree} (state~`\texttt{1}'). For additional technical details, we refer to %
\cite{heinrichsOpenSourceReconfigurable2023}.

\subsubsection{Wireless Environment and Devices}

We conduct all experiments in a typical office environment, where we deploy wireless devices representing the entities Alice, Bob, and Eve, as well as MTS-A and MTS-B. To accurately characterize the signal propagation behavior and the low-level physical dynamics of metasurface battles, we employ a \ac{VNA}. Specifically, we use a LibreVNA~\cite{LibreVNA2025} to measure the scattering parameter $S_{21}$—that is, the complex-valued frequency response—between two VERT2450 omnidirectional antennas placed approximately \SI{3}{\m} apart from each other.

For our system-level case studies, we use two USRP~N210s as controllable \mbox{Wi-Fi} transmitters, offering flexible signal generation and precise timing control, and a Raspberry~Pi~4B as a lightweight, commodity receiver representative of real-world client devices. Using the nexmon framework~\cite{gringoliFreeYourCSI2019}, the Pi’s onboard \mbox{Wi-Fi} interface is configured in monitor mode to capture fine-grained \ac{CSI} from received packets.

\begin{figure}
    \centering
    \includegraphics[width=0.9\linewidth]{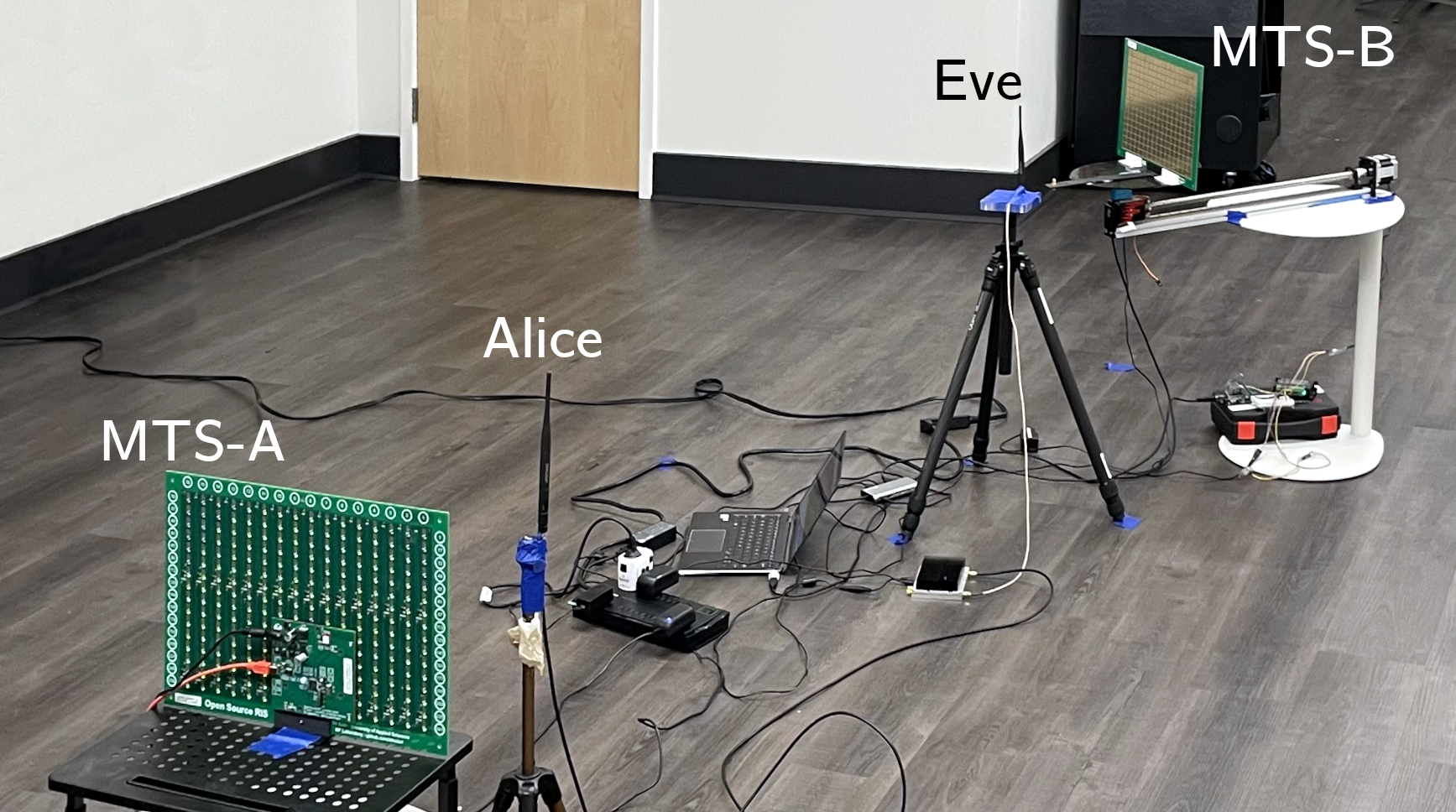}
    \caption{Experimental setup including both metasurfaces, antennas, measurement systems and the host computer.}
    \label{fig:experimental_setup_photo}
\end{figure}

All devices -- the metasurfaces and radio transceivers -- are orchestrated via a central control computer, which also collects and synchronizes all measurement data. The experimental setup is depicted in~\autoref{fig:experimental_setup_photo}.

\section{Case Study 1: Wireless Jamming}
\label{sec:jamming_demo}

\begin{figure}
    \begin{minipage}[c]{0.35\columnwidth}
        \begin{subfigure}{1\linewidth}
            \centering
            \includegraphics[width=1\linewidth]{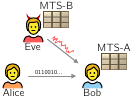}
            \caption{}
            \label{fig:cs_jamming}
        \end{subfigure}%
    \end{minipage}
    \hfill
    \begin{minipage}[c]{0.65\columnwidth}
        \begin{subfigure}{1\linewidth}
            \centering
            \includegraphics[width=1\linewidth]{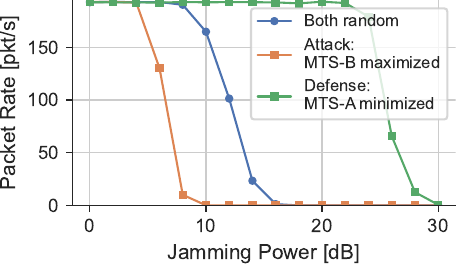}
            \caption{}
            \label{fig:cs_jamming_result}
        \end{subfigure}
        \end{minipage}\\ 
    \caption{Case study of a metasurface-battle about wireless jamming. (a)~Illustration of the involved parties, (b)~Bob's reception rate of packets from Alice over Eve's jamming power gain.}
\end{figure}

We begin our exploration of metasurface battles with a real-world example: \emph{metasurface-assisted wireless jamming}. Prior research has demonstrated that attackers can exploit metasurfaces to both actively and passively disrupt victim wireless communication~\cite{mackensenSpatialDomainWirelessJamming2025, staatMirrorMirrorWall2022, lyuIRSBasedWirelessJamming2020}. Conversely, metasurfaces can also serve as defensive tools to mitigate the impact of jamming~\cite{yaoAntiJammingTechniqueIRS2024}. In this study, we bring both perspectives together to form a complete metasurface battle, illustrated in~\autoref{fig:cs_jamming}. 

Inspired by~\cite{mackensenSpatialDomainWirelessJamming2025}, we consider a scenario where Eve employs MTS-B to improve their jamming channel toward the victim receiver Bob. In response, Bob deploys their own metasurface~MTS-A to defend against the interference. The resulting signal received by Bob, expressed in the equivalent complex baseband domain, is given by:
\begin{equation}
    Y = X H_{AB} + J H_{EB} + N,
\end{equation}
where $X$ denotes the desired \mbox{Wi-Fi} signal from Alice, $H_{AB}$ is the legitimate channel from Alice to Bob, $J$ represents the jamming signal transmitted by Eve, $H_{EB}$ is the corresponding jamming channel, and $N$ is additive white Gaussian noise. As illustrated in~\autoref{fig:system_model}, $H_{EB}$ is influenced by the configurations of both metasurfaces, MTS-A and MTS-B.

\Paragraph{Experiment Implementation.}
As the legitimate communication system, we use the USRP N210 \ac{SDR} to send IEEE~802.11n \mbox{Wi-Fi} packets (\SI{20}{\MHz} bandwidth with MCS value~0) from Alice to Bob, where they are received by a Raspberry~Pi~4B. The \mbox{Wi-Fi} chipset reports the \ac{RSSI} values and MAC~addresses of all successfully received packets. Eve's goal is to prevent Bob from correctly receiving \mbox{Wi-Fi} packets from Alice by sending their jamming signal towards Bob. For this, Eve employs a second USRP N210 \ac{SDR}, flooding Bob with malicious \mbox{Wi-Fi} packets. Eve places MTS-B next to their antenna while Bob places MTS-A next to the Raspberry Pi. 

Following~\cite{mackensenSpatialDomainWirelessJamming2025}, both metasurfaces employ a greedy optimization algorithm~\cite{tewesIRSenabledBreathTracking2022} based on \ac{RSSI} feedback: Eve tunes MTS-B to \emph{maximize} $|H_{EB}|$, while Bob tunes MTS-A to \emph{minimize} it. Eve estimates their jamming channel by eavesdropping on \mbox{Wi-Fi} packets from Bob, whereas Bob estimates Eve’s interference by monitoring the strength of Eve's jamming packets.

\Paragraph{Experiment Procedure and Results.}
Alice continuously transmits \mbox{Wi-Fi} packets at an approximate rate of 200~packets per second, and Bob’s packet reception rate serves as a direct indicator of link quality—and, inversely, of Eve’s jamming effectiveness. We first measure baseline performance with both metasurfaces randomly configured. As shown in~\autoref{fig:cs_jamming_result}, Eve requires roughly~\SI{16}{dB} of signal gain to fully suppress packet reception at Bob. When Eve optimizes MTS-B while Bob’s MTS-A remains random, their attack improves by approximately~\SI{6}{dB}, confirming the expected gain from metasurface-assisted jamming. To counteract the attack, Bob then optimizes MTS-A while MTS-B remains in the previously determined jamming configuration. As we can see from~\autoref{fig:cs_jamming_result}, the required jamming power for Eve to achieve full denial of service increases by about~\SI{20}{dB} compared to the attack state -- an improvement that not only neutralizes Eve’s metasurface gain but also significantly surpasses the non-battle baseline.

In this first demonstration of a metasurface battle, MTS-A clearly prevails, significantly reducing Eve’s attack performance beyond the initial non-battle baseline. This result illustrates that metasurface interactions are inherently dynamic and competitive rather than static or one-sided. Once both parties can adapt their configurations, dominance is no longer guaranteed for either side. This observation raises a fundamental research question that we will address in the next sections:
\begin{rqbox}[Research Question]
What are the key factors that determine which metasurface will prevail in a metasurface battle?
\end{rqbox}

\section{Analytical Model for Metasurface Battles}
\label{sec:analytical}

In the previous section, we saw an example for the most fundamental class of metasurface battles --- those centered on signal power. The amount of power reaching a receiver is the physical prerequisite for the correct operation of any wireless information system, making power control the most general and essential objective in both communication and sensing contexts. As demonstrated before, the defensive metasurface~MTS-A effectively suppressed the impact of the attacker’s metasurface~MTS-B, illustrating a competitive interaction over received signal power. This scenario represents one particular instance of a \emph{power battle}—a conflict in which multiple metasurfaces seek to manipulate the same propagation environment to their respective advantage.

At its core, such a battle can be viewed as a game between two metasurfaces that both influence the magnitude of a shared effective channel, here denoted as~$H_{EB}$. Each metasurface independently adapts its reflection states to either increase or decrease~$|H_{EB}|$, depending on its objective. This mutual dependency creates an inherently adversarial setting: improvements for one party directly degrade performance for the other.

\Paragraph{Gain Analysis.} Our analytical model (Appendix~\ref{sec:appendix}) captures the core mechanisms governing these interactions. A metasurface’s influence depends on the number of elements and the strength of the sub-channels connecting them to the antennas. Constructive surfaces amplify the channel by phase-aligning with existing propagation, while canceling surfaces reduce it by opposing the direct channel. Phase and amplitude mismatches limit both maximization and minimization, with minimizers being more constrained by the need for precise phase alignment. As a result, a minimizer is often more effective unless the maximizer’s path is substantially stronger. In fact, we show that if channels for both surfaces are equivalent, the gain of the battle favors the minimizer (see Appendix~\ref{sec:appendix} for the detailed proof):

\noindent \textbf{\textit{Lemma 1: }} 
The effective channel power gain of two equivalent metasurfaces with equivalent channels—one configured to maximize and the other to minimize the signal—can be approximated as $\left| 1 + r_{\text{max}} \left( e^{j \phi_{e_A}} - e^{j \phi_{e_B}} \right) \right|^2$ where $r_{\text{max}}$ denotes the metasurfaces' power relative to the non-metasurface channel. The outcome of this “battle” depends on which surface achieves smaller phase error ($\phi_{e_A}$ vs. $\phi_{e_B}$), which is typically the minimizer due to tighter phase-alignment requirements. \hfill $\square$

We further show that adversarial reconfigurations perturb the optimized channel, producing normally distributed amplitude and phase errors whose magnitude scales with the channel variance of the adversarial metasurface.

\Paragraph{Mutual \ac{SNR}.}
When both metasurfaces apply random configurations, each surface observes a superposition of its own contribution and the perturbations caused by the opponent, which can be modeled as additive noise: $c^A \mapsto |H_d + H_{MA} + N_B|$ and $c^B \mapsto |H_d + H_{MB} + N_A|$, with $N_B \sim \mathcal{CN}(0, \sigma_B^2)$ and $N_A \sim \mathcal{CN}(0, \sigma_A^2)$. The ability of a surface to reliably map its configurations to the observed channel depends on the effective \ac{SNR} of its contribution relative to the perturbations caused by the other surface. A lower \ac{SNR} makes it harder for the inferior surface to resolve which configuration improves or cancels the channel, reducing its effectiveness. In this sense, the mutual \ac{SNR} quantifies how distinguishable each surface’s effect is under interference from the other, and can serve as a predictor of which surface will dominate the channel.

\section{Factors Governing Power Battles}
\label{sec:factors}

To systematically understand what determines the outcome of power battles, we investigate three key factor dimensions:
($i$)~\emph{algorithmic behavior}, capturing how the metasurfaces explore and optimize their configurations;
($ii$)~\emph{channel effects}, including spatial placement, illumination, and environmental characteristics; and
($iii$)~\emph{hardware capabilities}, such as the number of metasurface elements.

In the following, we present results from a structured experimental study in which we systematically vary each of the battle dimensions. %
In all experiments, we use the LibreVNA to obtain high-quality measurements of the wireless channel between two antennas. Both metasurfaces have access to these channel measurements. This assumption reflects the most general and powerful attacker/defender model: by granting full channel knowledge, we consider worst-case scenarios in which each party can optimally adapt its configuration to maximize its advantage, thereby enabling a rigorous evaluation of the metasurface battle dynamics. Depending on their respective objectives, each surface attempts either to maximize or minimize the channel.

\subsection{Algorithms and Strategy}
\label{sec:algorithms}

A critical factor that determines both the nature and the outcome of a metasurface battle is the strategy each opponent employs. Battles may be prepared independently, respond adaptively, or unfold simultaneously. The latter represents the most intricate case, where both metasurfaces operate concurrently and continuously influence one another. As we will see, the outcome in such settings depends strongly on the particular optimization algorithm in use and the relative adaptation speed.

\begin{figure}
    \centering
    \includegraphics[width=1\linewidth]{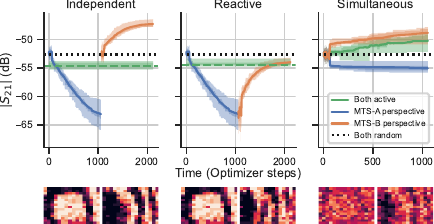}
    \caption{Evolution of wireless channel magnitudes during metasurface optimization considering three relative battle timings~(top). Resulting metasurface configurations~(bottom).}
    \label{fig:timing}
\end{figure}

\Paragraph{Relative Timing.}
In a metasurface battle, relative timing affects the outcome. We evaluated three different timings in an experiment where MTS-A and MTS-B again employ the greedy optimization algorithm to minimize and maximize the channel. The results are shown in~\autoref{fig:timing} where we plot the evolution of the channel magnitude $|S_{21}|$ for each surface and the resulting configurations of MTS-A and MTS-B.

\begin{enumerate}
\item \textbf{Independent optimization:} MTS-A is first optimized while MTS-B is configured static and random, reducing the baseline channel magnitude by approximately \SI{10}{\dB}. MTS-B then optimizes with MTS-A being static and random, achieving a gain of about \SI{6}{\dB}. When both surfaces apply their configurations simultaneously, the final channel settles roughly \SI{2}{\dB} below baseline, giving MTS-A the advantage.
\item \textbf{Reactive optimization:} MTS-A is optimized first and then remains in this state, allowing MTS-B to react to the already shaped channel. The resulting configuration of MTS-B only slightly differs from the independent optimization result, suggesting that MTS-A and MTS-B are largely independent of each other. In this mode, the final channel improves compared to independent optimization, remaining approximately \SI{1.5}{\dB} below baseline. 
\item \textbf{Simultaneous optimization:} Both MTS-A and MTS-B optimize in parallel, continuously adapting to each other. Here, each optimizer naively assumes that the channel is only influenced by itself, being unaware of the adversarial influence within the channel observations that guide the optimization process. For example, MTS-A falsely believes it can reduce the channel by \SI{2.5}{\dB}, but the actual channel (green line) rises approximately \SI{2.5}{\dB} above the baseline, giving MTS-B the upper hand. The resulting configurations indicate that MTS-B converges toward a stable state, while MTS-A appears more random, showing that simultaneous adaptation can prevent proper convergence of one party.
\end{enumerate}

These results demonstrate that the timing of optimization directly determines which metasurface gains the advantage. Independent or reactive sequences favor MTS-A, while simultaneous adaptation shifts the outcome toward MTS-B, highlighting the crucial role of timing in metasurface battles.

\Paragraph{Optimization Strategy.}
In the next experiment, we compare different optimization strategies in a simultaneous metasurface battle. Each party attempts to optimize the channel power using one of the following algorithms: a greedy genetic algorithm (GD)~\cite{tewesIRSenabledBreathTracking2022}, element-wise on–off flip testing (FL)~\cite{kainaShapingComplexMicrowave2014}, precomputed beamforming configurations (BF)~\cite{nayeriReflectarrayAntennasTheory2018}, linear regression (LR), purely random sampling (RD), and no optimization at all (NO).

\begin{figure*}
\centering
\begin{subfigure}{0.28\textwidth}
    \centering
    \includegraphics[width=1\linewidth]{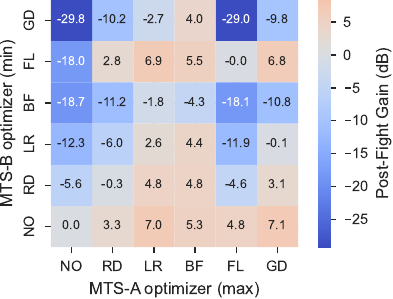}
    \caption{}
    \label{fig:algorithm_battle}
\end{subfigure}%
\hfill
\begin{subfigure}{0.28\textwidth}
    \centering
    \includegraphics[width=1\linewidth]{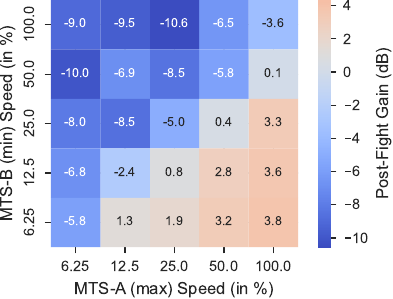}
    \caption{}
    \label{fig:speed}
\end{subfigure}%
\hfill
\begin{subfigure}{0.28\textwidth}
    \centering
    \includegraphics[width=1\linewidth]{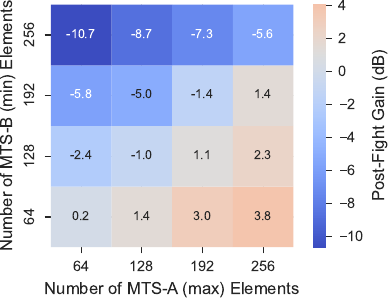}
    \caption{}
    \label{fig:number_of_elements}
\end{subfigure}
\caption{Channel gains relative to the random baseline after simultaneous battles between combinations of different: (a)~optimization algorithms, (b)~relative optimization speeds, (c)~active metasurface elements.}
\end{figure*}

\autoref{fig:algorithm_battle} shows the post-battle channel gains (relative to the random baseline) for all combinations of optimization strategies. We first assess each algorithm’s standalone effectiveness by comparing it against an opponent employing no optimization. In this non-adversarial setting, all algorithms succeed in either minimizing or maximizing the channel, with random sampling performing worst and the greedy algorithm achieving the strongest gains.

When both metasurfaces actively optimize, however, the interaction between algorithms substantially alters the outcome. A notable observation is the inferiority of the element-wise on–off flipping algorithm: while effective in isolation, it performs poorly in adversarial settings. Because FL modifies only a single metasurface element per iteration, its effect is easily overshadowed by algorithms that adjust multiple elements simultaneously, resulting in an asymmetric adaptation dynamic. Another interesting finding is the superiority of the beamforming-based algorithm (BF). By sampling random beam directions derived from one known antenna position and using corresponding precomputed metasurface configurations, BF gains an initial information advantage. This allows it to achieve strong early channel improvements which might mislead the adversarial iterative optimizers that rely solely on feedback-based updates.

Overall, the results show that both the greedy (GD) and beamforming (BF) strategies emerge as the most effective in direct competition, consistently outperforming other optimization methods across battle scenarios.

\vspace*{0.1in}\banner{The robustness of metasurface optimization strategies under adversarial conditions varies significantly across algorithms.}

\Paragraph{Speed.} 
We identified optimization speed as another decisive factor in metasurface battles. To study its impact, we conducted a simultaneous battle between MTS-A and MTS-B, both employing the greedy optimizer. We varied the relative optimizer speed by introducing configurable pauses: during these pauses, the respective optimizer remains idle, \ie, no new metasurface configuration is applied or evaluated. Each optimizer was paused for $1, 2, 4, 8,$ or $16$ steps, corresponding to relative speed ratios ranging from \SI{6.25}{\percent} to \SI{100}{\percent}.

\autoref{fig:speed} presents the resulting battle outcomes relative to the random baseline channel. Along the diagonal—where both optimizers operate at the same speed—MTS-B, which seeks to minimize the channel, maintains an advantage, consistently yielding a negative channel gain. However, as soon as MTS-A’s optimization speed exceeds that of MTS-B, MTS-A reliably gains the upper hand, with post-battle channel magnitudes rising above the baseline. These results underline that relative adaptation speed can decisively shift the balance of power in a metasurface battle, even when both sides employ identical algorithms.

\vspace*{0.1in}\banner{The faster metasurface is likely to win a battle.}

\subsection{Metasurface Capabilities}
\label{sec:metasurface_hw}

Another fundamental factor determining the outcome of a metasurface battle is the intrinsic capability of the metasurface hardware. Intuitively, a larger aperture allows a metasurface to intercept more of the incident electromagnetic energy and re-radiate it in controlled directions, thereby exerting greater influence over the propagation environment. However, this ability is not solely a function of aperture size: it also depends on the metasurface’s spatial and phase resolution. The spatial resolution, governed by the density of unit cells, dictates how finely the surface can approximate a desired reflection pattern, while the phase resolution controls how precisely each unit cell can tune its reflection coefficient. Coarse quantization in either domain limits the fidelity of wavefront synthesis, weakening the metasurface’s ability to form narrow or accurately directed beams. In addition, practical metasurfaces introduce insertion losses through their elements and control circuitry, leading to partial power dissipation. Together, these factors—aperture size, spatial and phase quantization, and intrinsic losses—define the upper bound of a metasurface’s control over the wireless channel and, consequently, its potential to prevail in a metasurface battle.

Since the physical aperture and quantization levels of our prototypes are fixed, we focus here on the effective number of active elements as a proxy for metasurface capability. To emulate reduced-size surfaces, we deactivate randomly selected unit cells by fixing them to random values, thereby limiting the number of elements available for optimization. Using this approach, we re-run the simultaneous metasurface battle while systematically varying the number of active elements on MTS-A and MTS-B, testing every combination.

The resulting post-battle channel gains are shown in~\autoref{fig:number_of_elements}. The heatmap reveals a clear pattern: the metasurface with more active elements consistently dominates the battle and reaches its respective optimization region. Moreover, the magnitude of the winning gain scales with the disparity in element counts, confirming that hardware capability directly translates into competitive advantage in metasurface battles.

\vspace*{0.1in}\banner{The metasurface with more elements has an edge in a battle, but this is not the sole determinant of success.}

\subsection{Channel Contributions}
\label{sec:channel_contributions}

For a metasurface to meaningfully influence the wireless channel, it must be sufficiently illuminated—that is, the sub-channels to and from its unit-cell elements must be strong enough to affect the overall propagation path. The degree to which a metasurface can assert dominance in a battle therefore depends on the structure of the underlying wireless channel. Three aspects are particularly critical:
($i$)~the randomness of the channel, including multipath fading effects;
($ii$)~the placement of the metasurface relative to the transmitting and receiving antennas, which determines path loss; and ($iii$)~the visibility of the metasurface, that is, how directly it participates in the dominant propagation paths.

\Paragraph{Channel conditions.}
To isolate the effect of varying channel conditions, we conduct the following experiment: MTS-A is positioned adjacent to Alice’s antenna and MTS-B next to Bob’s antenna, maintaining identical distances and geometry throughout. To vary the wireless channels between the antennas and the metasurfaces, we systematically sweep the carrier frequency. Although the metasurface response itself exhibits some frequency dependence, this effect is identical for the two technically identical devices (MTS-A and MTS-B), ensuring a fair comparison.

\begin{figure}
    \centering
    \includegraphics[width=0.8\linewidth]{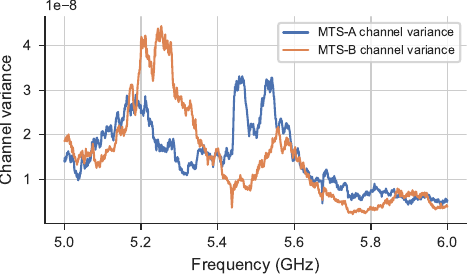}
    \caption{Variance of the channel magnitude due to random configuration of MTS-A and MTS-B over frequency.}
    \label{fig:frequency_channel_variation}
\end{figure}

For each frequency, we apply random configurations to both metasurfaces and measure the resulting variation of the wireless channel to quantify their respective influence. The results, shown in~\autoref{fig:frequency_channel_variation}, reveal that while both metasurfaces exhibit comparable overall impact, the magnitude of their effect fluctuates strongly with frequency. As the frequency changes, the underlying multipath conditions—and hence the illumination of each metasurface—vary randomly beyond the coherence bandwidth. Consequently, some frequencies favor MTS-A, while others favor MTS-B. We therefore conclude that even when all other parameters are balanced, random channel conditions can decide which metasurface dominates the battle.

\Paragraph{Location} 
The position of each metasurface relative to the antennas of Alice and Eve is a decisive factor in determining its influence—and therefore the outcome—of a battle. This dependence arises from the distance-related path loss of the channels to and from each metasurface, denoted $h^{m}_l$ and $g^{m}_l$ in~\autoref{eq:mts_subchannels}.

\begin{figure}
\centering
    \begin{subfigure}{1\linewidth}
        \centering
        \includegraphics[width=1\linewidth]{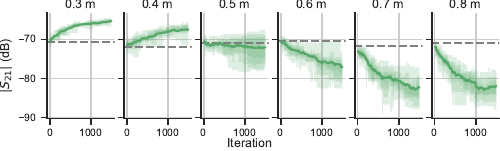}    
        \caption{}
        \label{fig:simult_evo_vs_distance}
    \end{subfigure}%
    \\
    \begin{subfigure}{1\linewidth}
        \centering
        \includegraphics[width=1\linewidth]{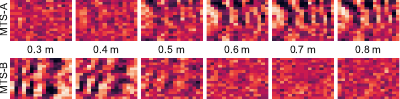}
        \caption{}
        \label{fig:simult_cfgs_vs_distance}
    \end{subfigure}
    \caption{Effect of varying the distance of MTS-B (maximizing) to the antenna in a simultaneous battle with MTS-A (0.3~m distance, minimizing).} \label{fig:simult_vs_dist}
\end{figure}

To quantify this effect, we perform a simultaneous metasurface battle in which MTS-A seeks to minimize and MTS-B to maximize the channel. MTS-A is fixed at a distance of \SI{0.3}{\m} from Alice’s antenna, while MTS-B is gradually moved from \SI{0.3}{\m} to \SI{0.8}{\m} away from Eve’s antenna. The resulting battle dynamics and final metasurface configurations are shown in~\autoref{fig:simult_vs_dist}. As the results illustrate, MTS-B is successful to maximize the channel and thus wins the battle when it is positioned less than \SI{0.5}{\m} from Eve's antenna. From \SI{0.5}{\m} on, the battle outcome starts to reverse and MTS-A now consistently is winning the battle, as the final channel gain is positive. This confirms that path-loss asymmetry directly translates into competitive advantage in metasurface battles.

\begin{figure}
\centering
    \begin{subfigure}{0.5\linewidth}
        \centering
        \includegraphics[width=0.95\linewidth]{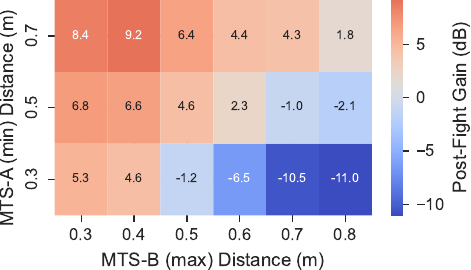}
        \caption{}
        \label{fig:joint_distance_post_fight}
    \end{subfigure}%
    \hfill
    \begin{subfigure}{0.5\linewidth}
        \centering
        \includegraphics[width=0.95\linewidth]{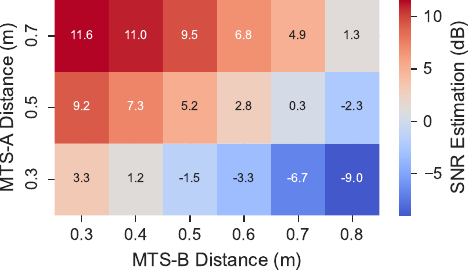}       
        \caption{}
        \label{fig:joint_distance_snr}
    \end{subfigure}%
    \caption{Battle dynamics under joint metasurface-antenna distance variation: (a)~Channel gain after simultaneous. (b)~\ac{SNR} estimation for MTS-B.}
    \label{fig:joint_distance}
\end{figure}

As we can see from \autoref{fig:joint_distance_post_fight}, this relationship is jointly determined by the distances of both MTS-A and MTS-B. As MTS-A moves further away from Alice' antenna, Eve can either realize higher post-battle gains or likewise move further away while still winning the battle.

Complementing our previous analytical model from~\autoref{sec:analytical}, we would like to use the effect of distance to show that the outcome of simultaneous metasurface battles aligns well with the mutual \ac{SNR} of MTS-A and MTS-B. First, we record the channels resulting from a sequence of random metasurface configurations on each surface, $s(c^A_i)$ and $s(c^B_i)$. Then, we apply the same random sequences again to both surfaces in parallel and record the resulting channel sequence $s(c^A_i, c^B_i)$. Using cross-correlation (matched filter detection), we estimate the signal energies of $\rho_A$ and $\rho_B$ of the individual sequences within the joint sequence. In the context of the previous example, given that the channel variation from MTS-A is considered as uncorrelated noise, we can estimate the \ac{SNR} of MTS-B from the ratio of $\rho_B$ and $\rho_A$, which we plot in \autoref{fig:joint_distance_snr}. This plot reveals a key finding: The joint distance variation of both surfaces determines the mutual \ac{SNR} which closely aligns with the resulting channel gains after the simultaneous battle shown in~\autoref{fig:joint_distance_post_fight}.

\begin{figure}
\centering
    \begin{subfigure}{1\linewidth}
        \centering
        \includegraphics[width=1\linewidth]{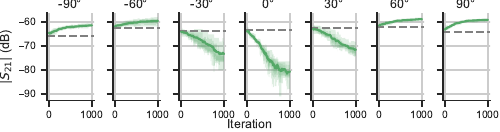}    
        \caption{}
        \label{fig:simult_evo_vs_angle}
    \end{subfigure}%
    \\
    \begin{subfigure}{1\linewidth}
        \centering
        \includegraphics[width=1\linewidth]{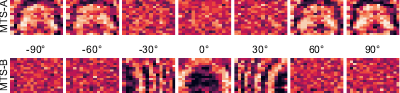}
        \caption{}
        \label{fig:simult_cfgs_vs_angle}
    \end{subfigure}
    \caption{Effect of varying the orientation of MTS-B (minimizing) to the antenna in a simultaneous battle with MTS-A (maximizing).} \label{fig:simult_vs_orientation}
\end{figure}

\Paragraph{Illumination and Visibility.}
In the next experiment, we investigate the effect of metasurface orientation on the battle outcome. To this end, we rotate MTS-B around its vertical axis using a motorized turntable by an angle~$\phi$ relative to the line-of-sight between Alice and Eve. This rotation scales the effective aperture of MTS-B, and thus its visible surface area, by approximately $\cos(\phi)$. As shown in~\autoref{fig:simult_vs_orientation}, for orientations within $\pm$~\SI{30}{\degree}, MTS-B wins the battle as it succeeds to minimize the channel. However, as $\phi$ increases and the MTS-B becomes less visible in the channel, its influence diminishes, allowing MTS-A to dominate the battle.

\vspace*{0.1in}\banner{Multipath variations, path-loss differences, and orientation-dependent visibility all shape metasurface interactions, contributing to—but not solely deciding—the outcome of a battle.}

\subsection{On Linearity Of Metasurface Battles}
\label{sec:metasurface_linearity}

An intriguing observation from our metasurface battle experiments is the emergence of non-linear coupling effects between the two surfaces, which often favor the minimizing party. Specifically, we found that the influence of one metasurface on the effective wireless channel can vary depending on the configuration of the other, indicating that the surfaces are not strictly independent.

To examine this effect, we conduct independent optimizations of MTS-A and MTS-B to obtain configurations $c^A$ and $c^B$ that define $H_{MA}$ and $H_{MB}$, respectively. Using ensemble averaging over the respective other surfaces to remove their effect, we then estimate the individual contributions in~\autoref{eq:system_model_basic}:
\begin{align}
    \hat{H}_A &\approx H_d + H_{MA}\\
    \hat{H}_B &\approx H_d + H_{MB}\\
    \hat{H}_d &\approx H_d\\
    \hat{H}_{\text{eff}} &\approx \hat{H}_A + \hat{H}_B - \hat{H}_d
\end{align}
If both surfaces act independently ($H_{MA,MB} \approx 0$), $\hat{H}_{\text{eff}}$ should approximate the measured $H_{\text{eff}}$ by superposition.

\begin{figure}
    \centering
    \includegraphics[width=0.7\linewidth]{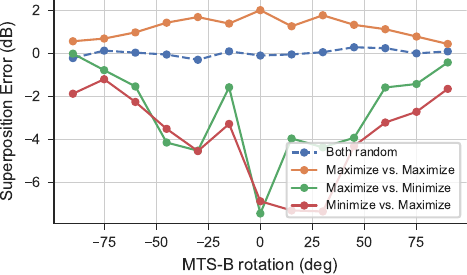}
    \caption{Effect of metasurface coupling on the effective channel magnitude under mutual optimization as a function of the metasurface angle.}
    \label{fig:superposition_vs_angle}
\end{figure}

However, as shown in~\autoref{fig:superposition_vs_angle}, the measured channel deviates significantly from this prediction when MTS-B faces MTS-A. In such configurations, coupling between the metasurfaces causes $H_{\text{eff}}$ to diverge from the superposed estimate—typically in favor of the minimizing party. When both surfaces maximize instead, the coupling becomes constructive, further increasing the channel magnitude. This deviation diminishes as MTS-B rotates away from MTS-A, where coupling weakens and superposition again holds. Please note that this effect does not occur for purely random configurations, where the superposition error remains around \SI{0}{\dB}.

\begin{figure}
    \centering
    \includegraphics[width=1\linewidth]{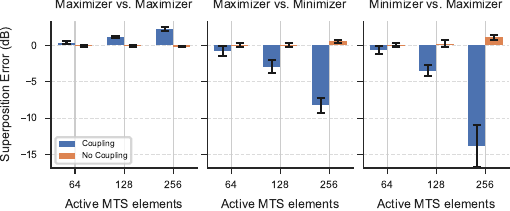}
    \caption{Effect of metasurface coupling on the effective channel magnitude under mutual optimization as a function of the metasurface elements.}
    \label{fig:comparison_barplot}
\end{figure}

As illustrated in~\autoref{fig:comparison_barplot}, this effect also scales with the number of active surface elements. In coupled configurations, deviations between the measured and expected channel magnitudes grow significantly, whereas in non-coupled settings they remain minimal. We thus conclude that metasurface coupling can be a non-negligible and asymmetric factor in metasurface battles—often benefiting the minimizer—but that it remains spatially constrained due to the high path loss inherent to dual-surface propagation.

\section{Generalization of Metasurface Battles}
\label{sec:casestudies}

While power control represents the most fundamental form of metasurface interaction, wireless systems exhibit far richer and more complex objectives. Metasurfaces can influence not only how much power reaches a receiver, but also \emph{where}, \emph{when}, and \emph{what} information is conveyed. To understand how the concept of metasurface battles extends to these broader contexts, we now explore three additional representative case studies spanning security and privacy applications.

Each case exemplifies a distinct dimension of adversarial metasurface interaction:
($i$)~directional misinformation, where metasurfaces compete to shape the spatial distribution of signals (\textsc{Protego}~\cite{liProtegoSecuringWireless2022});
($ii$)~~sensing obfuscation, where metasurfaces attempt to conceal or reveal physical movements (\textsc{IRShield}~\cite{staatIRShieldCountermeasureAdversarial2022}); and
($iii$)~~sensing manipulation, where metasurfaces generate misleading observations in active sensing systems (\textsc{RISiren}~\cite{jiangRISirenWirelessSensing2024}). 

Together, these case studies generalize the notion of metasurface battles beyond pure power competition, illustrating that strategic interactions can emerge across a diverse range of wireless security and privacy mechanisms.

\begin{figure*}
\centering
    \begin{subfigure}{0.2\textwidth}
        \centering
        \includegraphics[width=\linewidth]{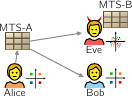}
        \caption{}
        \label{fig:cs_protego}
    \end{subfigure}%
    \hfill
    \begin{subfigure}{0.2\textwidth}
        \centering
        \includegraphics[width=\linewidth]{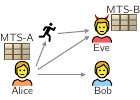}
        \caption{}
        \label{fig:cs_sens_obfusc}
    \end{subfigure}%
    \hfill
    \begin{subfigure}{0.2\textwidth}
        \centering
        \includegraphics[width=\linewidth]{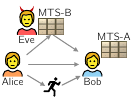}
        \caption{}
        \label{fig:cs_sens_spoof}
    \end{subfigure}
    \caption{Illustration of additional case study scenarios: (a) \textsc{Protego}: physical-layer secure communication~\cite{liProtegoSecuringWireless2022}, (b)~\textsc{IRShield}: sensing obfuscation for privacy~\cite{staatIRShieldCountermeasureAdversarial2022}, (c)~\textsc{RISiren}:~spoofing of sensing events~\cite{jiangRISirenWirelessSensing2024}.}
    \label{fig:sensitivity_types}
\end{figure*}

\subsection{Case Study 2: PHY Secure Communication}
\label{sec:protego}

We now turn to the \textsc{Protego} system~\cite{liProtegoSecuringWireless2022}, a metasurface-based approach for securing physical-layer (PHY) wireless communication between Alice and Bob through spatial channel obfuscation. \textsc{Protego} dynamically configures a metasurface to maintain a stable and coherent channel for legitimate communication while disrupting potential eavesdroppers through randomized phase perturbations. It achieves this via metasurface-assisted beamforming that concentrates reflected signal energy toward Bob while injecting controlled randomness into other spatial directions. The overall scenario is illustrated in~\autoref{fig:cs_protego}.

\Paragraph{Experiment Implementation.}
Alice uses a directional antenna aimed at MTS-A, which implements \textsc{Protego}. Bob and Eve are positioned at distinct angles relative to the metasurface, with Eve deploying MTS-B next to their antenna. Alice transmits a \ac{QPSK}-modulated signal that reaches both Bob and Eve after reflection from MTS-A. To measure the complex-valued channels defined by \textsc{Protego}, we employ the LibreVNA with an external \ac{RF} switch alternating between the antennas of Bob and Eve. Using the recorded channel responses, we simulate the transmission of $1000$ \ac{QPSK} symbols from Alice to both receivers.

\begin{figure}
    \centering
    \includegraphics[width=1\linewidth]{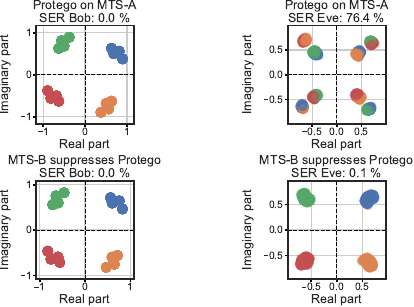}
    \caption{Received \ac{QPSK} constellations for Bob (left) and Eve (right) with \textsc{Protego} active, shown without (top) and with (bottom) Eve's metasurface MTS-B optimized.}
    \label{fig:protego_modulation}
\end{figure}

\Paragraph{Experiment Procedure and Results.}
We first evaluate \textsc{Protego} in a non-battle setting. Here, the adversarial metasurface MTS-B remains randomly configured, while MTS-A applies the \textsc{Protego} scheme, which selects configurations from a discrete set designed to obfuscate Eve’s channel, see~\cite{liProtegoSecuringWireless2022} for additional details. The resulting \ac{QPSK} constellation diagrams of Bob and Eve are shown in the top row of \autoref{fig:protego_modulation}. Bob observes a clean, coherent constellation with clearly distinguishable constellation points, yielding a \ac{SER} of \SI{0}{\percent}. In contrast, Eve’s received symbols are randomly phase-shifted into false quadrants, resulting in an \ac{SER} of \SI{76.4}{\percent}, effectively matching the \SI{75}{\percent} error rate expected from random guessing in a four-symbol alphabet. This confirms the intended behavior of \textsc{Protego}: the legitimate channel remains stable, while the eavesdropper’s channel becomes decorrelated and unusable.

We now extend this setup into a full metasurface battle. Eve actively deploys MTS-B beside their antenna to counteract \textsc{Protego}’s obfuscation. Using the greedy optimization algorithm introduced in~\autoref{sec:algorithms}, Eve adapts MTS-B to minimize the circular standard deviation of the observed channel phases—effectively suppressing the spatial randomness imposed by MTS-A. After optimization, we repeat the \ac{QPSK} transmission experiment and analyze the resulting constellations, shown in the bottom row of~\autoref{fig:protego_modulation}. Bob’s link remains unaffected, maintaining an \ac{SER} of \SI{0}{\percent}, confirming that MTS-B does not interfere with the legitimate communication path. In stark contrast, Eve’s decoding performance improves dramatically: their \ac{SER} drops to just \SI{0.1}{\percent}. In other words, Eve successfully neutralizes \textsc{Protego}, restoring channel coherence and recovering the transmitted symbols with near-perfect accuracy. 

To understand why this attack succeeds, consider how Eve receives the signals. The \textsc{Protego}-modulated transmissions from MTS-A reach Eve directly, but they also impinge on MTS-B, which reflects an additional, phase-shifted copy of the same signal back toward Eve. By tuning its reflection state, MTS-B effectively synthesizes a compensating signal that cancels the random phase perturbations introduced by \textsc{Protego}. As a result, the artificial temporal variation intended to obfuscate Eve’s channel is neutralized at their antenna, restoring a coherent signal. In the terminology of our battle framework, the evaluation function~$\mathcal{G}$ can here be interpreted as Eve’s \ac{SER} performance—showing that MTS-B (Eve) decisively wins the metasurface battle against MTS-A running \textsc{Protego}.

This result illustrates how the defensive advantage of a metasurface can be challenged by an equally capable adversary: While \textsc{Protego} can protect the legitimate channel against passive eavesdropping, it can be defeated by an adaptive, metasurface-equipped attacker.

\vspace*{0.1in}\banner{Metasurfaces that disrupt eavesdroppers through random phase perturbation can have their obfuscation undone by a metasurface-equipped adversary.}

\subsection{Case Study 3: IRShield -- Sensing Obfuscation}
\label{sec:irshield}

In our next case study, we examine the \textsc{IRShield} channel obfuscation scheme~\cite{staatIRShieldCountermeasureAdversarial2022}, designed to protect users against adversarial wireless sensing. The concept, illustrated in~\autoref{fig:cs_sens_obfusc}, deploys a metasurface next to a \mbox{Wi-Fi} router to introduce randomness into the wireless channel of potential eavesdroppers. Specifically, \textsc{IRShield} continuously reconfigures the metasurface using randomized reflection states and periodic configuration inversions. This dynamic behavior injects artificial variations into the observed channel, thereby preventing the adversary from reliably detecting human motion or other environmental changes—effectively enhancing privacy in indoor environments.

\Paragraph{Experiment Implementation.}
To emulate the \textsc{IRShield} scenario, we position a USRP~N210~\ac{SDR} within a closed room (Alice) that periodically transmits \mbox{Wi-Fi} packets, functioning as a household access point. A metasurface (MTS-A) operating the \textsc{IRShield} algorithm is placed adjacent to Alice’s antenna. Outside the room, Eve uses a Raspberry~Pi~4B in monitor mode to eavesdrop the \mbox{Wi-Fi} signals and extract \ac{CSI} from each received packet. Eve’s objective is to detect human motion inside the room using a simple variance-based detector that computes a sliding standard deviation over the observed \ac{CSI} data~\cite{zhuTuAlexaWhen2020}. Challenging \textsc{IRShield}, Eve places their own metasurface (MTS-B) next to their Raspberry Pi receiver, attempting to stabilize the channel and counteract the obfuscation introduced by MTS-A.

\begin{figure}
\begin{subfigure}{1\columnwidth}
    \centering
    \includegraphics[width=1\linewidth]{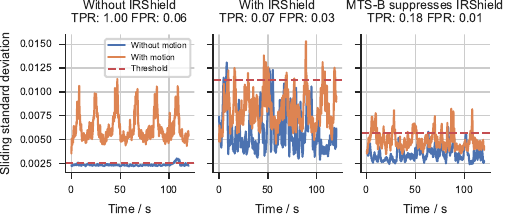}
    \caption{}
    \label{fig:irshield_strong}
\end{subfigure}\\
\begin{subfigure}{1\columnwidth}
    \centering
    \includegraphics[width=1\linewidth]{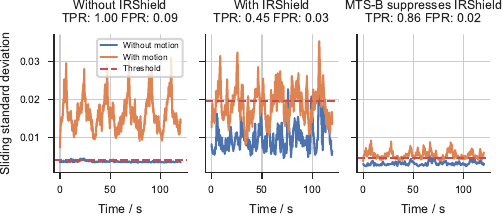}
    \caption{}
    \label{fig:irshield_weak}
\end{subfigure}
\caption{Adversarial human motion detection using eavesdropped \ac{CSI}. Left: attack baseline, middle: MTS-A applies the \textsc{IRShield} defense, right: attack based on optimization of MTS-B. Top: MTS-A directly facing towards the adversary. Bottom: MTS-A facing away from the adversary.}
\end{figure}

\Paragraph{Experiment Procedure and Results.}
We evaluate the \textsc{IRShield} scenario in three stages, comparing the attacker’s ability to detect human motion within the victim environment. As a baseline, both metasurfaces MTS-A and MTS-B are configured randomly and remain static, representing the case without any \textsc{IRShield} protection. In this setting, a person continuously walks in circles within the room for approximately two minutes to generate repeatable motion patterns. As we can see from~\autoref{fig:irshield_strong}~(left), Eve is clearly able to distinguish between the static environment and human motion based on the eavesdropped \ac{CSI} data, demonstrating that movement within the room can be reliably detected.

Next, we activate \textsc{IRShield} on MTS-A, enabling continuous randomization of the metasurface configuration while measuring both static and motion scenarios under the same configuration sequence. Under these conditions, the distinction between human motion and static states becomes largely obscured by the artificial channel fluctuations introduced by MTS-A, as shown in~\autoref{fig:irshield_strong}~(middle). As a result, Eve’s motion-detection accuracy drops sharply, demonstrating that \textsc{IRShield} effectively conceals physical movement and disrupts adversarial wireless sensing.

Finally, we examine the metasurface battle in which Eve actively attempts to counter \textsc{IRShield} by using the greedy optimization algorithm introduced in~\autoref{sec:algorithms} to find a metasurface configuration that minimizes temporal channel variation. To ensure a fair comparison, we reuse the same sequence of \textsc{IRShield} configurations as before. During optimization, however, Eve observes a separate, randomly generated set of \textsc{IRShield} configurations—reflecting the inherently random nature of \textsc{IRShield} and preventing overfitting to a specific temporal channel pattern. As shown in~\autoref{fig:irshield_strong}~(right), Eve is able to partially restore motion-detection capability but cannot completely overcome the obfuscation, as the distributions for motion and no-motion conditions still overlap significantly.

Compared to the earlier attack on \textsc{Protego}, this setting presents a more complex optimization challenge for Eve: \textsc{Protego} employs a small, predefined set of four crafted configurations, whereas \textsc{IRShield} uses continuous randomization with no inherent limit on the configuration space. Nevertheless, we found that the attacker can still defeat \textsc{IRShield} under less favorable deployment conditions. As shown in~\autoref{fig:irshield_weak}, when MTS-A is oriented away from the adversary—reducing the obfuscation effect perceived by Eve—the protection weakens. In this case, even without optimization of MTS-B, Eve can already distinguish motion more clearly than before. Once MTS-B is optimized, the adversary fully restores motion-detection performance, achieving an almost perfect separation between static and motion states.

\vspace*{0.1in}\banner{Metasurfaces can defeat sensing obfuscation from another metasurface, although the battle outcome depends on the orientation of the surfaces.}

\subsection{Case Study 4: RISiren -- Sensing Manipulation}
\label{sec:risiren}

In the \textsc{RISiren} attack~\cite{jiangRISirenWirelessSensing2024}, the adversary, Eve, employs a malicious metasurface to manipulate the wireless propagation channel between Alice and Bob, thereby spoofing wireless sensing events. The attack concept is illustrated in~\autoref{fig:cs_sens_spoof}. Specifically, the \textsc{RISiren} metasurface alternates between two states, introducing controlled temporal variations into the wireless channel that mimic Doppler-frequency signatures corresponding to (fake) physical activities within the environment.

\Paragraph{Experiment Implementation.}
The experimental setup mirrors the configuration used in previous case studies. Alice uses a USRP~N210~\ac{SDR} to transmit periodic \mbox{Wi-Fi} packets, while Bob, equipped with a Raspberry Pi~4B operating in monitor mode, receives the packets and records the corresponding \ac{CSI}. Bob’s objective is to infer human activities based on \ac{CSI} time-series analysis. To this end, we implement the Falldefi framework~\cite{palipanaFallDeFiUbiquitousFall2018}, following the methodology in~\cite{jiangRISirenWirelessSensing2024}. Falldefi applies a short-time Fourier transform (\ac{STFT}) to convert the temporal channel variations into Doppler-frequency spectrograms, which serve as input to an activity classification model.

To launch the attack, Eve positions their metasurface (MTS-B) adjacent to Alice’s transmitter. In response, Bob deploys a defensive metasurface (MTS-A) near their receiver, consistent with the metasurface battle notion.

\begin{figure}
    \begin{subfigure}{1\columnwidth}
        \centering
        \includegraphics[width=1\linewidth]{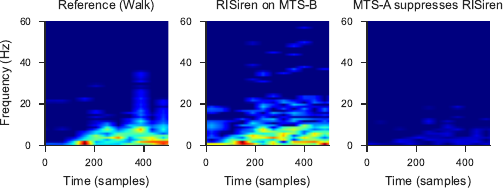}
        \caption{}
        \label{fig:risiren_walk}
    \end{subfigure}\\
    \begin{subfigure}{1\columnwidth}
        \centering
        \includegraphics[width=1\linewidth]{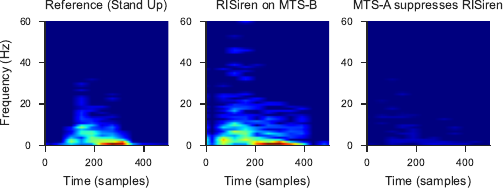}
        \caption{}
        \label{fig:risiren_standup}       
    \end{subfigure}
    \caption{Doppler frequency spectra for human activity recognition of walking~(top) and stand up~(bottom) events. Left: Real human activity, middle: MTS-B conducts \textsc{RISiren}-based channel spoofing, right: MTS-A suppresses the \textsc{RISiren} spoofing.}
    \label{fig:risiren_results}
\end{figure}

\Paragraph{Experiment Procedure and Results.}
We begin by recording spectrograms corresponding to two distinct human activities—walking and standing up—which serve as reference events. As shown in the left column of~\autoref{fig:risiren_results}, the two activities produce clearly distinguishable Doppler signatures. Using these reference spectrograms, Eve employs a genetic optimization algorithm to synthesize a switching sequence for MTS-B. This sequence is designed to produce channel variations that, when processed by Falldefi, mimic the Doppler patterns of the reference activities.

Following the procedure outlined in~\cite{jiangRISirenWirelessSensing2024}, Eve exploits knowledge of the relative positions of Alice and Bob to compute two optimized MTS-B configurations that respectively maximize and minimize the channel gain between them. By toggling between these two configurations according to the synthesized sequence, Eve induces artificial channel dynamics. Despite the absence of any real movement in the environment, the resulting Doppler spectrograms (middle column of~\autoref{fig:risiren_results}) closely replicate those of genuine human activities, confirming the efficacy of the \textsc{RISiren} spoofing mechanism.

Next, we turn the \textsc{RISiren} attack scenario into a metasurface battle: Eve’s MTS-B seeks to manipulate the wireless channel to inject spoofed sensing signatures, while Bob’s MTS-A aims to suppress these artificial perturbations and preserve sensing integrity. Following the assumptions of~\cite{jiangRISirenWirelessSensing2024}, Eve continuously operates MTS-B to sustain the illusion of human activity. Because the \textsc{RISiren} channel variations are abrupt and discrete -- unlike the smoother fluctuations caused by genuine human activity -- Bob can detect the attack and leverage MTS-A to defend. Using the greedy optimization algorithm introduced in~\autoref{sec:algorithms}, we minimize the variance in \ac{CSI} induced by MTS-B’s switching behavior. As we can see from the left column in~\autoref{fig:risiren_results}, Bob successfully cancels out the effect of MTS-B, effectively suppressing Eve's spoofing attempts and thereby restoring sensing capabilities.

\vspace*{0.1in}\banner{Metasurfaces can be used to sense and mitigate wireless sensing attacks from another metasurface.}

\section{Related Work}
In addition to the references on unilateral metasurface applications in the security and privacy domain already mentioned in the Introduction~\cite{staatMirrorMirrorWall2022, mackensenSpatialDomainWirelessJamming2025, shaikhanovMetaFlyWirelessBackhaul2024, weiMetasurfaceenabledSmartWireless2023, liProtegoSecuringWireless2022, xuTwodimensionalHighorderDirectional2024, shaikhanovSpoofingEavesdroppersAudio2025, staatIRShieldCountermeasureAdversarial2022, zhouRIStealthPracticalCovert2023, jiangRISirenWirelessSensing2024, reddyvennamMmSpoofResilientSpoofing2023, chenMetaWaveAttackingMmWave2023, wangSpreadSpectrumSelectiveCamouflaging2021, shaoTargetMountedIntelligentReflecting2024}, a recent survey of Wang~\etal~\cite{wangNavigatingDualUseNature2025} explicitly discusses the benign and adversarial applications of metasurfaces in a security context, yet without considering a direct \enquote{battle}. To the best of our knowledge, the only contributions on metasurfaces with conflicting objectives are of Alexandropoulos~\etal~\cite{alexandropoulosCounteractingEavesdropperAttacks2023} and Rezvani~\etal~\cite{rezvaniLegitimateIllegitimateIRSs2022}, which contribute secrecy rate analyses in the context of information-theoretic code-based physical-layer security. In contrast, in our work, we consider the general case of metasurfaces with conflicting objectives which we complement with an extensive experimental evaluation.

Strategic behavior of parties with conflicting objectives in a physical-layer context has been studied with game-theoretic frameworks for adversarial wireless sensing \cite{wijewardenaPlugnPlayGameTheoretic2020} and jamming~\cite{zhangDynamicAntiJammingStrategy2024, jiaGameTheoryReinforcement2025}, and \acp{RIS}-assisted information-theoretic physical-layer security \cite{zhaiImprovingPhysicalLayer2022, luongDynamicNetworkService2021}. Despite these advances, we are, to the best of our knowledge, the first to characterize and empirically evaluate the interaction of multiple metasurfaces with conflicting objectives.

\section{Conclusion, Discussion and Future Work}

This work has explored metasurface \enquote{battles} in various application contexts, demonstrating their profound implications for security and privacy at the wireless physical layer and highlighting numerous directions for future work.

\Paragraph{Extension to Multiparty Scenarios}
While this work focuses on metasurface battles between two adversarial entities, the concept naturally extends to multiparty settings involving more than two metasurfaces with potentially conflicting objectives. However, such scenarios introduce additional complexity: if the battle evaluation function~$\mathcal{G}$ is defined only for pairwise interactions, the presence of multiple metasurfaces implies the emergence of coalitions rather than purely adversarial relationships. In other words, when $\mathcal{G}$ remains binary while more than two metasurfaces are involved, some participants must implicitly collaborate—forming temporary alliances—to compete against common opponents. Extending the framework to genuinely multiparty battles therefore requires redefining $\mathcal{G}$ to capture higher-order strategic dependencies and coalition dynamics.

\Paragraph{Dynamic Behavior on Both Sides}
In this work, we have primarily focused on metasurface battles in which at least one opponent ultimately adopts a static configuration. Although we have examined cases where both metasurfaces exhibit dynamic behavior, the corresponding battle evaluation was constrained to scenarios in which one side eventually commits to a fixed configuration optimized against an opponent employing multiple configurations—two in~\autoref{sec:risiren}, four in~\autoref{sec:protego}, and an arbitrary number in~\autoref{sec:irshield}. We envision future studies extending this framework to fully dynamic battles, where both adversaries can use more than one configurations.

\Paragraph{Adaptation Strategies}
This paper identified key factors that influence the outcome of a metasurface battle. To simplify the analysis, we assumed that each party behaves naively, strictly following strategies derived from a non-adversarial wireless environment. This opens the door to exploring battle-adapted strategies that could balance trade-offs—such as maintaining robust performance in benign settings while achieving significant gains under adversarial conditions.

\Paragraph{Electromagnetic Cloning}
Beyond the metasurface battles explored in this work, a particularly interesting direction for future study is what we term \emph{electromagnetic cloning}. Unlike existing spoofing techniques that deceive sensing algorithms through artificial channel perturbations, electromagnetic cloning aspires to replicate the true physical scattering response of a real object or scene at the waveform level. In this setting, a metasurface would function as an electromagnetic twin, reproducing the spatial, spectral, and temporal field characteristics of its target. A natural extension of this idea is to investigate whether such electromagnetic clones could withstand scrutiny from an opposing metasurface explicitly configured to perform reality plausibility checks—that is, to validate whether observed fields are physically consistent with the surrounding environment. Exploring this adversarial interplay would provide deep insights into the fundamental limits of perception and deception in smart radio environments, revealing how close engineered wave control can come to reproducing, or distinguishing, electromagnetic reality itself.

\section*{Acknowledgments}

We thank NovoFlect GmbH, Simon Tewes, Markus Heinrichs, and Rainer Kronberger for providing the metasurface prototypes. We acknowledge support from the NSF (2106921, 2030154, 2433903, 2007786, 1942902, 2111751), ONR, AFRETEC, and CyLab-Enterprise. and the Deutsche Forschungsgemeinschaft~(DFG, German Research Foundation) under Germany’s Excellence Strategy - EXC 2092 CaSa - 390781972.

\section*{LLM usage considerations}
LLMs were used for editorial purposes in this manuscript, and all outputs were inspected by the authors to ensure accuracy and originality.

\bibliographystyle{IEEEtran}
\bibliography{meta_battle}

\appendices

\section{Analytical Model}
\label{sec:appendix}

The metasurfaces MTS-A and MTS-B have $L_A$ and $L_B$ unit-cell reflection elements with phase-programmable reflection coefficients $c_l = e^{j \phi_l}$. For simplicity, we consider the reflection elements to be loss-less, regardless of their configuration, \ie, $|c^A_l| = |c^B_l| = 1$. Thus, each element acts as an ideal phase shifter. We assume that MTS-A and MTS-B are functionally identical metasurfaces. That is, both surfaces match in size, arrangement, and number of unit-cell reflection elements. The reflection coefficients behave the same, implying that for identical surface configurations, MTS-A and MTS-B exhibit the same radiation behavior.
We abbreviate the combined illumination channels between the respective metasurface elements and the antennas as $h^{A}_l g^{A}_l = a_l$ and $h^{B}_l g^{B}_l = b_l$.

\subsection{Channel Randomization}
Assuming $L_m \gg 1$ and uniformly random chosen metasurface configurations, $H_{Mm}$ converges to a complex normal distribution, \ie, $H_{Mm} \sim \mathcal{CN}(0, \sigma_m^2)$. The channel variance is found by evaluating the expected value of the squared channel, where:
\begin{align}
    \sigma_m^2 &= \mathbb{E}[ H_{Mm}^2 ] - \mu_m^2\\
    &= \sum_l^{L_m} |h^m_l g^m_l|^2
\end{align}

Thus, we conclude that the effect of metasurface channel variation without optimization is governed by the channels towards and from the metasurface to the antennas and the number of elements.

\subsection{Gain analysis}
In the following, we assume that MTS-A seeks to maximize and MTS-B to minimize $|H_{\text{eff}}|$. Both metasurfaces are independently optimized while the respective other metasurface is static and randomly configured. Here, we ignore the impact of the dual-surface channel component $H_{MA,MB}$. Further, for the sake of simplicity, we assume that both metasurfaces are capable of continuous phase shifting.

\subsubsection{Maximization}
To maximize $|H_{\text{eff}}|$, MTS-A has to phase-align all $L_A$ sub-channels of $H_{MA}$ to ensure constructive interference. For loss-less reflection elements with $|c^A_l| = 1$, the maximum amplitude is
\begin{equation}
|H^{max}_{MA}| = \sum^{L_A}_l |a_l|.
\end{equation}
In case of uniform sub-channel amplitudes, $|a_l| = a$ for all $l$, the squared magnitude is $|H^{max}_{MA}|^2 = |L_A \cdot a|^2$, \ie, power scales with $L_A^2$.

To maximize $|H_{\text{eff}}|$, $H_{MA}$ must be phase-aligned with the remaining channel components that are not controlled by MTS-A, \ie, $H'_d = H_d + H_{MB}$, thus 
\begin{equation}
    H^{max}_{MA} = |H^{max}_{MA}| e^{j\phi'_d}.
\end{equation}

In this case, the amplitude of the effective channel becomes
\begin{equation}
    \max_{c_p} |H_{\text{eff}}| = |H'_d| + |H^{max}_{MA}|
\end{equation}
and the resulting power gain is
\begin{equation}
    G'_{max} = \frac{(|H'_d| + |H^{max}_{MA}|)^2}{|H'_d|^2}.
\end{equation}

In practical scenarios, the metasurface optimization will be imperfect to some extent. We capture such imperfection through an error phasor $\eta_A e^{j \phi_{e_A}}$. $\eta_A$ is an amplitude-efficiency factor $0 < \eta_{A} \leq 1$ and $\phi_{e_A}$ is the phase deviation of $H_{MA}$ from $arg(H'_d)$:
\begin{equation}
    H^{max'}_{MA} = \eta_A\ e^{\phi_{e_A}} H^{max}_{MA}
\end{equation}
In this case, the effective channel amplitude becomes:
\begin{align*}
    |H_{\text{eff}}| &= \left| |H'_d| e^{j \phi'_d} + \eta_A\ |H^{max}_{MA}| \ e^{j \phi'_d} e^{j \phi_{e_A}} \right| \\
    &= \left| |H'_d| + H^{max'}_{MA} \right|\\
\end{align*}

\subsubsection{Minimization}

Minimization of $|H_{\text{eff}}|$ demands forming a cancellation signal that destructively interferes with $H'_d = H_d + H_{MA}$. Thus, the metasurface MTS-B configuration $c^B$ needs to be chosen such that $H_{MB} = -H'_d$. %

We capture an imperfect cancellation signal through the error phasor $\eta_B e^{j \phi_{e_B}}$, describing relative amplitude and phase mismatches. Thus, the cancellation signal of MTS-B is denoted as
\begin{equation}
    H^{min'}_{B} = \eta_B |H_d'| e^{j(\phi_d' + \phi_{e_B} + \pi)},
\end{equation}
The resulting effective channel then is:
\begin{equation}
    H_{\text{eff}} = |H'_d|e^{j\phi_d'} + \eta_B |H_d'| e^{j(\phi_d' + \phi_{e_B} + \pi)},
\end{equation}
with the amplitude being
\begin{align*}
    |H_{\text{eff}}| &= |H'_d| \left| 1 - \eta_B e^{j\phi_{e_B}}\right|.
\end{align*}
This leads us to the power cancellation ratio
\begin{align*}
    G'_{min} = \frac{|H_{\text{eff}}|^2}{|H'_d|^2} &= \left| 1 - \eta_B e^{j\phi_{e_B}}\right|^2\\
    &= 1 - 2\eta_B \cos(\phi_{e_B}) + \eta_B^2
\end{align*}

In the error-free case, \ie, $\eta_B e^{j \phi_{e_B}} = 1$, the direct signal is negatively matched, fully canceling $|H'_d|$ therefore and $|H_{\text{eff}}| = 0$. For small amplitude and phase errors, \ie, $\eta_B = (1 - \epsilon) \approx 1$ and $\phi_{e_B} \approx 0$, we can approximate $G'_{min} \approx \phi_{e_B}^2 + \epsilon^2$.%

\subsection{Ideal Superposition}

MTS-A attempts to maximize the amplitude of $H_{MA}$ while phase-aligning it with $H_d$. MTS-B attempts to negatively match $H_{MB}$ to $H_d$. After we have considered individual surface optimization before, we now take a look at the resulting channel when both surfaces come together in a battle. We consider the direct channel $|H_d|$ as the reference to compare the effective dual-metasurface channel against. If $|H_{\text{eff}}| > |H_d|$, MTS-A wins. If $|H_{\text{eff}}| < |H_d|$, MTS-B wins.

The resulting combined channel is:
\begin{align*}
    H_{\text{eff}} &= H_d + H^{max'}_{MA} + H^{min'}_{MB}\\
    &= |H_d|e^{j \phi_d} \left( 1 + \frac{\eta_A\ |H^{max'}_{MA}|}{|H_d|} \ e^{j \phi_{e_A}} - \eta_B e^{j \phi_{e_B}} \right)
\end{align*}
Thus, the post-battle channel power gain is:
\begin{align}
    \frac{|H_{\text{eff}}|^2}{|H_d|^2} &= \left|  1 + \frac{|H^{max'}_{MA}|}{|H_d|} \ e^{j \phi_{e_A}} - \eta_B e^{j \phi_{e_B}} \right|^2\\
\end{align}
Which in the absence of phase- and amplitude-errors becomes
\begin{equation}
    \frac{|H_{\text{eff}}|^2}{|H_d|^2} = \frac{|H^{max}_{MA}|^2}{|H_d|^2},
\end{equation}
confirming the intuition that for the maximizer to win in the presence of a perfect minimizer, it must accomplish a signal amplitude stronger than the direct signal.

Assuming that the maximum channel amplitude from MTS-B cannot fully match the amplitude of $H_d$, $\eta_B < 1$, and we may rewrite:
\begin{equation}
    |H_{\text{eff}}| = |H_d| \left|  1 + \frac{|H^{max'}_{MA}|}{|H_d|} \ e^{j \phi_{e_A}} - \frac{|H^{max'}_{MB}|}{|H_d|} e^{j \phi_{e_B}} \right|.
\end{equation}
If we further assume that both surfaces enjoy the same channels $H^{max'}$, we arrive at:
\begin{equation}
    |H_{\text{eff}}| = |H_d| \left|  1 + \frac{|H^{max'}|}{|H_d|} \ \left(e^{j \phi_{e_A}} - e^{j \phi_{e_B}} \right) \right|.
\end{equation}
From this equation, we can see that the battle would end up in draw, if both metasurfaces achieve the same gains without any phase errors. However, we believe that it is more likely that the minimizer achieves a lower phase error (which is a necessity for amplitude-constrained minimization) while the maximizer is less constrained to this end. If we assume $\phi_{e_B} \approx 0$ and $\phi_{e_A}$ to be small:
\begin{align*}
    \frac{|H_{\text{eff}}|^2}{|H_d|^2} %
    &\approx 1 + \left(\frac{|H^{max'}|^2}{|H_d|^2} - \frac{|H^{max'}|}{|H_d|})\right)\ \phi_{e_A}
\end{align*}
Since we assume that $\frac{|H^{max'}|}{|H_d|} < 1$, the second term will be negative, indicating that the minimizer will win the battle.

\subsubsection{Optimization Degradation due to Adversarial Reconfiguration}

When a metasurface is optimized while another metasurface is in place, the direct channel considered for optimization is perturbed and the optimized settings incur error terms $\eta_A e^{j\phi_{e_A}}$ and $\eta_B e^{j\phi_{e_B}}$.
\begin{enumerate}
    \item %
    Each surface optimizes against a non-MTS channel that includes the current (possibly random) configuration of the other surface:
    \[
    H'_d = 
    \begin{cases}
        H_d + H_{MB}, & \text{when MTS-A optimizes},\\[3pt]
        H_d + H_{MA}, & \text{when MTS-B optimizes.}
    \end{cases}
    \]

    \item %
    \begin{itemize}
        \item \emph{MTS-A (maximizer):} aligns its sub-channels to match $\arg(H'_d)$, producing $H^{\text{max}}_{MA} = |H^{\text{max}}_{MA}| e^{j\phi'_d}$ to boost $|H_{\text{eff}}|$.
        \item \emph{MTS-B (minimizer):} aligns to oppose $\arg(H'_d)$, producing $H^{\text{min}}_{MB} \approx -H'_d$, i.e., $H^{\text{min}}_{MB} = \eta_B |H'_d| e^{j(\phi'_d + \pi)}$, to cancel $|H_{\text{eff}}|$.
    \end{itemize}

    \item %
    Because each surface tailors its setting to $H'_d$, the result is an optimization of the perturbed---not the true direct---channel $H_d$. This mismatch gives rise to the error terms $\eta_A e^{j\phi_{e_A}}$ and $\eta_B e^{j\phi_{e_B}}$.

    \item %
    If the other metasurface later changes its configuration, $H'_d$ changes and the previously optimized setting no longer matches the new channel, reducing constructive gain or destructive cancellation. During optimization, the direct channel may be written as $H'_{d,0} = H_d + e_0 $. After adversarial reconfiguration, the channel then changes to $H'_{d,1} = H_d + e_1$.

\end{enumerate}

\textbf{Effect on the Minimizer.}
The initial metasurface MTS-B optimization matches $H'_{d,0}$:
\begin{equation}
    H_{\text{eff}} = H'_{d,0} - \eta_B |H'_{d,0}| e^{j\phi'_{d,0}} e^{j\phi_{e_B}}
\end{equation}
After reconfiguration of the opposite metasurface MTS-A, the effective channel is:
\begin{equation}
    H'_{\text{eff}} = H'_{d,1} - \eta_B |H'_{d,0}| e^{j\phi'_{d,0}} e^{j\phi_{e_B}},
\end{equation}
such that optimized channel becomes outdated. We now absorb this change in an updated optimization error:
\begin{equation}
    H'_{\text{eff}} = H'_{d,1} - \eta'_B |H'_{d,1}| e^{j\phi'_{d,1}} e^{j\phi'_{e_B}}
\end{equation}
where
\begin{equation}
    \eta'_B e^{\phi_{e_B}} = \eta_B \frac{|H_{d,0}|}{|H_{d,1}|} e^{j(\phi_{d,0} - \phi_{d,1} + \phi_{e_B})}
\end{equation}
Thus, the amplitude error increases proportional to the amplitude ratio of the baseline channel. The phase error increases by the phase difference of the baseline channels.

\textbf{Effect on the Maximizer}
The initial metasurface MTS-A optimization matches $H'_{d,0}$:
\begin{equation}
    H_{\text{eff}} = H'_{d,0} + \eta_A |H^{max}_{MA}| e^{j\phi'_{d,0}} e^{j\phi_{e_A}}
\end{equation}
After reconfiguration of the opposite metasurface MTS-B, the effective channel is:
\begin{equation}
    H'_{\text{eff}} = H'_{d,1} + \eta_A |H^{max}_{MA}| e^{j\phi'_{d,0}} e^{j\phi_{e_A}},
\end{equation}
such that optimized channel becomes outdated. We again absorb the change of the baseline channel in an updated optimization error:
\begin{equation}
    H'_{\text{eff}} = H'_{d,1} + \eta'_A |H^{max}_{MA}| e^{j\phi'_{d,1}} e^{j\phi'_{e_A}}
\end{equation}
where
\begin{equation}
    \eta'_A e^{\phi_{e_A}} = \eta_A e^{j(\phi'_{d,0} - \phi'_{d,1} + \phi_{e_A})}
\end{equation}
Other than for the minimizer, the amplitude error is not coupled to baseline amplitude. Therefore, in this case, only the phase error increases by the phase difference of the baseline channels.

\textbf{Degradation under Random Configurations}
In this case, $e_0$ and $e_1$ correspond to the metasurface channels that are the result of applying two different random configurations. To understand how this affects the optimization quality, we view the change of $H'_{d,0}$ to $H'_{d,1}$ as an error term on the optimizer side and re-use the results from the previous section. Thus, we are interested in the amplitude and phase changes, $\eta$ and $\phi_e$, between these two channels:
\begin{equation}
    H_e = \frac{H'_{d,1}}{H'_{d,0}}
\end{equation}

We assume randomized metasurface configurations, such that $e_0, e_1 \in \mathcal{CN}(0, \sigma_m^2)$, causing relatively small perturbations of $H_d$, \ie, $\sigma \ll |H_d|$, allowing us to approximate 
\begin{equation}
    H_e \approx 1 + \frac{e_1 - e_0}{H_d}
\end{equation}
Since the second term is small, we may write:
\begin{equation}
    |H_e| \approx 1 + \Re \left\{ \frac{e_1 - e_0}{H_d} \right\}
\end{equation}
Since $\frac{e_1 - e_0}{H_d} \sim \mathcal{CN}(0, \frac{2\sigma_m^2}{|H_d|^2})$, it follows that $\Re \{ \frac{e_1 - e_0}{H_d} \} \sim \mathcal{N}(0, \frac{\sigma_m^2}{|H_d|^2})$, and thus $|H_e| \sim \mathcal{N}(1, \frac{\sigma_m^2}{|H_d|^2})$.

For the phase, we similarly can write
\begin{align}
    \phi_e &\approx \arg(1 + \frac{e_1 - e_0}{H_d})\\
    &\approx \Im \left\{\frac{e_1 - e_0}{H_d}\right\}
\end{align}
and thus, $\phi_e \sim \mathcal{N}(0, \frac{\sigma_m^2}{|H_d|^2})$.

After all, this result shows that the damage imposed on the optimized metasurface is maximized either when the direct channel is minimal or when maximizing the difference between the channels $e_1$ and $e_0$, which can be accomplished by fully inverting the metasurface configuration. In any case, given the variance $\sigma_m^2$ of $e_0$ and $e_1$, we again observe a link with the incident channels and the number of elements of the respective metasurface.

\end{document}